*Science Letters:*

# Three-dimensional conceptual model for service-oriented simulation[*]

Wen-guang WANG[†1], Wei-ping WANG[1], Justyna ZANDER[2], Yi-fan ZHU[1]

(*[1]College of Information System and Management, National University of Defense Technology, Changsha 410073, China*)
(*[2]Fraunhofer Institute FOKUS, Kaiserin-Augusta-Allee 31, Berlin 10589, Germany*)
[†]E-mail: wgwangnudt@gmail.com

**Abstract:**    In this letter, we propose a novel three-dimensional conceptual model for an emerging service-oriented simulation paradigm. The model can be used as a guideline or an analytic means to find the potential and possible future directions of the current simulation frameworks. In particular, the model inspects the crossover between the disciplines of modeling and simulation, service-orientation, and software/systems engineering. Finally, two specific simulation frameworks are studied as examples.

**doi:**10.1631/jzus.A0920258            **Document code:**  A            **CLC number:**  TP391.9

INTRODUCTION

With the widespread application of modeling and simulation (M&S) techniques, there is an emerging need to combine efforts from numerous domains. Service-Oriented Architecture (SOA) (Erl, 2005; Papazoglou and van den Heuvel, 2007) has been a key enabler to extend the capabilities of M&S frameworks such as interoperability, composability, extensibility, agility, and reusability (Chen, 2007) in net-centric environments.

In terms of simulation, various service-oriented simulation frameworks have been proposed or implemented by different institutes using different formalisms or techniques. These include formalism-based (Mittal, 2007), model-driven (Tsai *et al.*, 2006a), interoperability protocol based (Wang *et al.*, 2008), or Open Grid Services Architecture (OGSA) based (Li *et al.*, 2005) frameworks.

However, the frameworks proposed so far generally focus on specific domains or systems and each has its pros and cons. They are capable of addressing different issues within service-oriented simulation from different viewpoints. It is increasingly important to develop a high level conceptual model that can describe and evaluate the progress of numerous service-oriented simulation frameworks. Such a model would help to identify the potential and deficiencies of current frameworks and facilitate research, development, improvements and the application of the old and new frameworks. Finally, it would lead to new solutions to the reusability, composability and interoperability of heterogeneous simulation resources.

In this letter, interrelated concepts are explored first. Then, we propose a three-dimensional conceptual model as a conceptual guideline and analytical method for service-oriented simulation frameworks. We clarify its implications from the viewpoint of one, two, and three dimensions (1D, 2D, and 3D). Finally, two specific service-oriented simulation frameworks are studied as examples.

---
[*] Project (Nos. 60574056 and 60674069) supported by the National Natural Science Foundation of China

CONCEPT EXPLORATION

In the definitions of a 'service' given by (W3C, 2004; Balin and Giard, 2006; Quartela et al., 2006; DoD, 2007), a lot of attention is paid to capability, utility, interface, and functionality aspects whereas implementation details are generally hidden.

A 'simulation service' refers to the capability of M&S activities owned or implemented by abstract (i.e., conceptual) or concrete (i.e., implementation-related) elements that can be used by other services.

'Service-oriented simulation' is defined as a simulation using a service-oriented paradigm in which the service and its capability are considered more important than the object, process, or outcome, etc. It focuses on description, publication and composition in the lifecycle of services or simulation services.

A 'service-oriented M&S framework' is the fundamental organization of a service-oriented simulation system representing its components (services/simulation services), its components' relationships to each other and to the environment in the process of modeling, description, publication, composition, execution, etc., and the principles that guide its design and evolution.

THREE-DIMENSIONAL CONCEPTUAL MODEL FOR SERVICE-ORIENTED SIMULATION

**Overview of the 3D model**

Similar to the 3D morphology of systems engineering (Hall, 1969), research on service-oriented simulation involves (at least) three distinct, yet related fundamental dimensions (domains or viewpoints): M&S, service-orientation, and software/systems engineering. These three dimensions comprise the conceptual model of service-oriented simulation (Fig.1). The model represents the state-of-the-art and can be applied as an engineering reference model or a framework for the analysis, design, implementation, evaluation, and evolution of service-oriented simulation systems.

There are two directions in the service-oriented simulation domain (Chen, 2007). One is the use of SOA in M&S, i.e., employing a service-oriented paradigm to extend the capacity of M&S techniques and frameworks. An example is the design and implementation of simulators that are services themselves, and can be invoked via SOA protocols (Mittal, 2007). The other is a vice versa approach, where M&S is used for SOA, i.e., M&S techniques are applied to address the problems in service-oriented systems. An example is the application of simulators that evaluate models of software packages designed along the SOA paradigm (Tsai et al., 2006a; 2006b). Similarly, there are two levels in service-oriented simulation: the problem to be simulated, and the simulation mechanism. Both can be service oriented. The 3D model is intended to cover both directions and both levels by using different results that are Cartesian products from different orders of M&S and service-orientation dimensions.

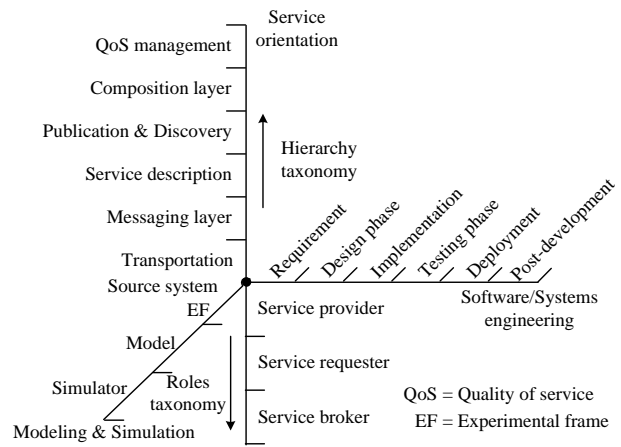

Fig.1 Three-dimensional conceptual model for service-oriented simulation

Regarding the number and layout of dimensions, there may exist multi-dimensions, sub-dimensions or negative dimensions (Waite, 2006). Service-oriented simulation must cover at least three dimensions such as M&S, service orientation, and engineering, as explained in the following subsections. In addition, other dimensions or sub-dimensions such as systems of systems (Jamshidi, 2009) and different levels of interoperability (Wang et al., 2009) exist. However, it is hard to imagine and understand issues generated beyond 3D. Thus, for simplicity, we do not include them here. Furthermore, any additional elements can be regarded as parts of the main three dimensions (e.g., systems of systems and levels of interoperability can be complementary to the engineering dimension). Moreover, the service orientation dimension is broken down along a positive and negative axis. Hence, we stick to the three dimensions depicted in Fig.1.

**1D implication**

A 1D view enables us to look at each fundamental dimension individually. The source system is located at the origin. It stands for the existing or proposed system that we intend to observe or test.

1. M&S dimension

Besides the source system, the basic entities in M&S (Zeigler *et al.*, 2000) include the model, simulator, and experimental frame (EF). Modeling and simulation are the fundamental relationships. Following the logical flow of general M&S practice, EF is firstly determined as the operational formulation of the M&S objectives. Then, we develop a model to represent the source system under a certain EF. Finally, we use a simulator to execute the model to generate its behavior for further study. Hence, we use the sequence of EF, model, and simulator in this dimension.

2. Service-orientation dimension

This dimension has two taxonomies that come from the conceptual structure of SOA and the implementation hierarchies of Web services, respectively. The two taxonomies are complementary and the combination of them can better facilitate the analysis and implementation of service-oriented applications.

One of the taxonomies, from the viewpoint of roles, is structured as a triangle that consists of a service provider, requester, and broker. We use this particular order for this scale because the service provider and requester are more fundamental roles than the service broker. The service provider must provide its service earlier than the requester's demand so as to compose a successful application.

The other taxonomy, from the perspective of Web service stack, is where the hierarchies of transportation, messaging, service description, service publication and discovery, composition and collaboration, and quality of service (QoS) management appear. Transportation, messaging and service description are the core layers that constitute the basis for static SOA. Service publication and discovery, composition and collaboration levels enhance the dynamic capabilities for dynamic SOA. QoS management makes services more dependable and robust by focusing on QoS requirements (Yu *et al.*, 2007) such as performance, reliability, scalability, interoperability, and security. We sequence the elements by their decreasing importance on the scale in Fig.1.

3. Software/systems engineering dimension

Simulation systems usually include software, at least in part (McKenzie *et al.*, 2004). The research on software engineering, especially architecture and lifecycle, are of great help to simulation systems. Systems engineering can also contribute to service-oriented simulation in terms of hardware, optimization, trade-off, or decision making aspects. The lifecycle of software/systems engineering may be assigned to different ontologies from multiple viewpoints (IEEE Std. 15288, 2008). In this letter, we use the taxonomy of requirement, design (e.g., description, design, analysis, etc.), implementation, testing, deployment, and post-development (e.g., maintenance, evolution, reuse, etc.). In fact, the activities along the engineering dimension are often cyclic or concurrent.

**2D implication**

A 2D view inspects each domain created by the Cartesian products of every two dimensions. It reveals the cross-discipline landscape of service-oriented simulation.

1. Narrow service-oriented simulation

The Cartesian product of M&S and service-orientation dimensions lets us treat the service-oriented simulation in a narrow sense. It has two implications that reveal the two directions of SOA for M&S and vice versa, respectively: an approach that enables an extension of the traditional M&S artifacts with the service-orientation principles, and an approach that models or simulates service-oriented systems by means of M&S. Similarly, as for the previous discussion, the Cartesian product of different dimensions provides different directions. It is the fundamental domain of service-oriented simulation. We call it a 'narrow approach' because it lacks rigorous engineering principles or processes.

2. M&S engineering

The Cartesian product of M&S and software/systems dimensions provides an M&S engineering domain. It applies engineering principles and methods to traditional M&S as in, for example, the classical High Level Architecture (HLA) Federation Development and Execution Process standards (IEEE Std. 1516.3, 2003). It is the traditional M&S engineering domain that does not necessarily refer to service-oriented simulation.

3. Service-oriented engineering

The Cartesian product of service-orientation and software/systems dimensions creates a service-oriented engineering domain. Here, engineering principles are applied to a service-orientation community. Although the basic engineering principles seem still unchanged (along the classical engineering dimension), new requirements and challenges are introduced by the SOA paradigm. For example, services are key elements, service interfaces, reuse and composition are paid more attention to, and the development style is mainly model driven. Service-oriented engineering is a new emerging domain. Typical examples include service-oriented systems engineering (Tsai, 2005) and service-oriented software engineering (Tsai *et al.*, 2008). In particular, these authors discussed the impact of the SOA paradigm on classical software/systems engineering principles and practices.

**3D implication**

A 3D view provides a complete multi-perspective consideration of a service-oriented simulation. The whole 3D space is constituted by the Cartesian product of all the dimensions. This represents 'service-oriented M&S engineering', also called 'general service-oriented simulation' because it applies engineering principles to the whole development lifecycle of service-oriented simulation systems. To evolve as a new and mature M&S paradigm, service- oriented simulation must cover the whole 3D space demanded by the 3D model.

We identify the core issues (if they are not present, the framework cannot be called a service-oriented simulation framework), supporting issues (if they are missing, the framework will be heavily impacted), and also nice-to-have issues (if they do not appear, the framework may be slightly impacted). These are listed in Tables 1~3. Also, the model and simulator belong to the core, whereas the experimental frame for nice-to-have issues is in the M&S dimension.

**Table 1  Application of the 3D model from model view**

|  | Broker | Requester | Provider | Transportation | Messaging | Description | P&D | Composition | QoS |
|---|---|---|---|---|---|---|---|---|---|
| Req |  |  | F1 |  | F1 | F1, F2 |  |  |  |
| Dsn |  | F1 (User), F2 (PSML$^T$) | F1 (compile, transform, validate), F2 (PSML$^T$) | F1, F2 (PSML$^T$) | F1, F2 (PSML$^T$) | F1 (DEVSML), F2 (PSML$^T$) |  | F1 (SES, static), F2 (PSML$^T$) | F2 |
| Imp |  | F1 (User), F2 (PSML$^T$) | F1 (compile, transform, validate), F2 (PSML$^T$) | F1, F2 (PSML$^T$) | F1, F2 (PSML$^T$) | F1 (DEVSML), F2 (PSML$^T$) |  | F1 (SES, static), F2 (PSML$^T$) | F2 |
| Tst |  |  | F1 |  |  | F1 |  |  | F2 |
| Dply |  |  | F1, F2 | F1, F2 | F1, F2 | F1, F2 |  | F2 | F2 |
| PstD |  |  | F2 | F2 | F2 | F2 |  | F2 | F2 |

The increasing gray intensity of the cells identifies nice-to-have, supporting, and core issues, respectively. Elements marked with a superscript 'T' (transposition) identify M&S for SOA; normal elements identify SOA for M&S. F1=Discrete Event System Specification (DEVS) Unified Process (DUNIP) framework; F2=Dynamic Distributed Service-oriented Simulation (DDSOS) framework; SES=System Entity Structure; DEVSML=Discrete Event System Specification Modeling Language; Req=Requirement; Dsn=Design; Imp=Implementation; Tst=Test; Dply=Deploy; PstD=Post-development; P&D=Publication & Discovery

**Table 2  Application of the 3D model from simulator view**

|  | Broker | Requester | Provider | Transportation | Messaging | Description | P&D | Composition | QoS |
|---|---|---|---|---|---|---|---|---|---|
| Req |  |  |  |  | F1 |  |  |  |  |
| Dsn | F2 | F1 (User), F2 | F1, F2 | F1, F2 | F1, F2 | F1, F2 | F2 | F2 | F2 |
| Imp | F2 | F1 (User), F2 | F1, F2 | F1, F2 | F1, F2 | F1, F2 | F2 | F2 | F2 |
| Tst |  |  |  |  |  |  |  |  |  |
| Dply |  |  | F1, F2 | F1, F2 | F1, F2 | F1, F2 |  | F2 | F2 |
| PstD |  |  | F2 | F2 | F2 | F2 |  | F2 | F2 |

The increasing gray intensity of the cells identifies nice-to-have, supporting, and core issues, respectively. F1=Discrete Event System Specification (DEVS) Unified Process (DUNIP) framework; F2=Dynamic Distributed Service-oriented Simulation (DDSOS) framework; Req=Requirement; Dsn=Design; Imp=Implementation; Tst=Test; Dply=Deploy; PstD=Post-development; P&D=Publication & Discovery

Table 3  Application of the 3D model from experimental frame view

|      | Broker | Requester | Provider | Transportation | Messaging | Description | P&D | Composition | QoS |
|------|--------|-----------|----------|----------------|-----------|-------------|-----|-------------|-----|
| Req  |        |           |          |                |           | F1          |     |             |     |
| Dsn  |        |           | F2       | F1, F2         | F1, F2    | F1, F2      |     | F2          | F2  |
| Imp  |        |           | F2       | F1, F2         | F1, F2    | F1, F2      |     | F2          | F2  |
| Tst  |        |           |          |                |           |             |     |             |     |
| Dply |        |           |          | F1             | F1        | F1          |     |             |     |
| PstD |        |           |          |                |           |             |     |             |     |

The gray intensity of the cells identifies nice-to-have issues. F1=Discrete Event System Specification (DEVS) Unified Process (DUNIP) framework; F2=Dynamic Distributed Service-oriented Simulation (DDSOS) framework; Req=Requirement; Dsn=Design; Imp=Implementation; Tst=Test; Dply=Deploy; PstD=Post-development; P&D=Publication & Discovery

The 3D conceptual model can be applied to separate concerns and used as a taxonomy of the existing service-oriented simulation frameworks. More-over, it can aid domain experts to define clearer and more specific activities. It can also help discover new research issues for multiple discipline experts so that sub-phases or steps can then be added using the Cartesian products. Examples of possible research problems generated by crossing the service-orientation and M&S dimensions include how to capsulate the capability of models, simulators, and experimental frames as services, and how to manage, use, and implement them at respective layers. From the engineering point of view, the properties, design, and implementation problems should be considered as complements to the above issues.

APPLICATION OF THE 3D MODEL

We take two specific service-oriented simulation frameworks as examples to show the usefulness of the 3D model. One is the Discrete Event System Specification (DEVS) Unified Process (DUNIP) framework (Mittal, 2007), and the other is the Dynamic Distributed Service-Oriented Simulation (DDSOS) framework (Tsai *et al.*, 2006a; 2006b).

**DUNIP service-oriented simulation framework**

DUNIP belongs to formalism-based frameworks. It can automatically generate DEVS models from a number of different specifications. DEVS models are regarded as resources, while simulators are distributed Web services. DEVS Modeling Language (DEVSML) (Mittal *et al.*, 2007a) was proposed to represent DEVS models with XML format. DEVS/SOA (Mittal *et al.*, 2007b; Jamshidi, 2009) was proposed for model execution over the net-centric infrastructure.

We put the DUNIP framework into the 3D space as shown in Tables 1~3. Applying the descriptive role of the 3D model, we can see that the DUNIP covers the core area and also some of the supporting areas. Employing the prescriptive role of the 3D model, the broker, publication and discovery, composition, and QoS are not incorporated, and the engineering process can be further enhanced.

**DDSOS service-oriented simulation framework**

DDSOS belongs to model-driven frameworks. Process Specification and Modeling Language for Services (PSML-S) (Tsai *et al.*, 2007a) was proposed for service-oriented systems. The mappings from SOA artifacts to PSML elements were reported in (Tsai *et al.*, 2007b). HLA Runtime Infrastructure (RTI) with some extensions is taken as the simulation infrastructure. DDSOS has distinct functionalities such as dynamic simulation federation configuration management, automated simulation code generation and deployment, as well as multi-agent simulation for reconfiguration and dynamic analysis.

We analyzed the DDSOS framework based on the same set of tables as before. Applying the descriptive role of the 3D model, we can see that the DDSOS covers some of the cells, in particular, modeling SOA artifacts, dynamic composition, and post-development. Based on the prescriptive role of the 3D model, the broker [e.g., dynamic brokers (Tsai *et al.*, 2007c)], and publication & discovery are missing, while the PSML can be further improved and standardized to be compatible with other formalisms and applicable for further communities.

**Summary**

The descriptive and analytic functionalities of the 3D model lead to several outcomes. DUNIP and DDSOS currently represent the emphasis of SOA on M&S and vice versa, respectively, though they have some potential to cover both directions. DUNIP has the advantages of a rigorous theoretical basis, compatibility with many other specifications, and many extensions. However, recent research (Tsai *et al.*, 2009) shows that it is mainly an object-oriented approach. In contrast, DDSOS has other advantages. It covers more issues in the service-oriented dimension. It allows for simulation of a service-oriented approach by PSML. Still, the theoretical basis, standardization, and practices of PSML can be further enhanced. To sum up, there is still room to improve the specific frameworks so that the full spectrum of capabilities depicted in the 3D model can be covered.

CONCLUSION

We have proposed a 3D conceptual model for service-oriented simulation frameworks. It reveals the landscape of the research issues in a service-oriented simulation domain. It can be used as a guideline or as a method of analysis to find the potential and possible future directions of current service-oriented simulation frameworks. In future work, we aim to improve the model by addressing more issues generated from both directions of SOA for M&S and vice versa. It will also be used to evaluate, compare, and describe various current classical service-oriented simulation frameworks. It will be a key enabler to make service- oriented simulation a new simulation paradigm following on from object- and process-oriented methods.

ACKNOWLEDGEMENTS

The authors thank the anonymous reviewers for their constructive comments. The authors also thank Prof. Pieter J. Mosterman and Dr. Yinong Chen for the valuable suggestions.

---

Welcome to reading or referring to the authors' publications:

[1] Wang W.G., Wang W.P., Zander J., Zhu Y.F., 2009. Three-dimensional conceptual model for service-oriented simulation. *Journal of Zhejiang University SCIENCE A*, **10**(8):1075-1081.
http://www.zju.edu.cn/jzus/2009/A0908/A090801.htm
http://www.springerlink.com/content/p217u140771hx35q/?p=281e44773f0b47d09220f5d992b913c9&pi=2

[2] Wang W.G., Xu Y.P., Chen X., Li Q., Wang W.P., 2009. High level architecture evolved modular federation object model. *Journal of Systems Engineering and Electronics*, **20**(3):625-635. http://arxiv.org/abs/0909.1364

[3] Wang W.G., Tolk A., Wang W.P., 2009. The levels of conceptual interoperability model: Applying systems engineering principles to M&S. *Spring Simulation Multiconference (SpringSim'09)*. San Diego, CA, USA. http://arxiv.org/abs/0908.0191

[4] Wang W.G., Yu W.G., Li Q., Wang W.P., Liu X.C., 2008. Service-oriented high level architecture. *European Simulation Interoperability Workshop*. 08E-SIW-022. http://arxiv.org/abs/0907.3983

Best kind regards

Wenguang Wang
(wgwangnudt@gmail.com)